\documentclass[aps,twocolumn,showpacs, showkeys,prb]{revtex4}
\usepackage{graphicx}
\usepackage{graphicx,color,epsfig,rotate}
\usepackage{dcolumn}
\usepackage{bm}
\input{epsf.sty}

\begin{document}
\newcommand\hcone{$H_{c1}$}
\newcommand\hctwo{$H_{c2}$}
\newcommand\bacusio{BaCuSi$_2$O$_6$}

\title{Magnetostriction in the Bose-Einstein Condensate quantum magnet NiCl$_2$-4SC(NH$_2$)$_2$}
\author{V. S. Zapf$^1$, V. F. Correa,$^{2,{\dagger}}$ C. D. Batista,$^3$ T. P. Murphy$^2$, E. D. Palm$^2$, M. Jaime,$^1$ S. Tozer$^2$, A. Lacerda$^1$, A.
Paduan-Filho$^4$ \\$^1$National High Magnetic Field Laboratory
(NHMFL), Los
Alamos National Lab (LANL), Los Alamos, NM \\
$^2$NHMFL, Tallahassee, Florida \\
$^3$Condensed Matter and Thermal Physics, LANL, Los Alamos, NM \\
$^4$ Instituto de Fisica, Universidade de Sao Paulo, Brazil \\
$^{\dagger}$ Now at Comisi\'on Nacional de Energ\'ia At\'omica,
Centro At\'omico Bariloche, 8400 S. C. de Bariloche, Argentina}

\date{\today}

\begin{abstract}

The quantum magnet NiCl$_2$-4SC(NH$_2$)$_2$ is a candidate for
observing Bose-Einstein Condensation of spin degrees of freedom in
applied magnetic fields. An XY antiferromagnetic ordered state
occurs in a dome-shaped region of the temperature-field phase
diagram between H$_{c1}$ = 2.1 T and H$_{c2}$ = 12.6 T and below
1.2 K. BEC corresponds to the field-induced quantum phase
transition into the ordered state. We investigate magnetostriction
in single crystals of this compound at dilution refrigerator
temperatures in magnetic fields up to 18 T, and as a function of
magnetic field angle. We show that significant changes in the
lattice parameters are induced by magnetic fields, and argue that
these result from antiferromagnetic couplings between the Ni spins
along the tetragonal c-axis. The magnetic phase diagram as a
function of temperature, field, and field angle can be extracted
from these data. We discuss the implications of these results to
Bose-Einstein Condensation in this system.

\end{abstract}

\maketitle

\section{Introduction}

In recent years, there has been a surge of interest in the topic
of quantum magnetism in condensed-matter systems. Nature and human
ingenuity have provided us with a rich variety of spin structures
including reduced-dimensional ladders, chains or planes, dimers,
frustrated spins, and single-molecule magnets. These materials all
share the common trait that quantum effects, such as spin
fluctuations and quantized spin levels, play a significant role in
shaping the ground state and the physical properties of the
system. The resultant behavior can be complex and challenge our
understanding, as well as provide potentially useful applications.

In particular, there have been a number of attempts recently to
observe the quintessential quantum ground state Bose-Einstein
Condensation (BEC) in quantum magnets. BEC was first observed in
dilute gases of $^{87}$Rb atoms, \cite{Anderson95} leading to a
nobel prize being awarded in 2001. It turns out that a form of BEC
can also be observed in quantum magnets, e.g. crystalline lattices
containing spins. BEC in quantum magnets was first predicted to
occur in 1991 by Ian Affleck \cite{Affleck91}. In the past few
years, several reports of BEC in real spin systems have been
published including TlCuCl$_3$,\cite{Nikuni00,Yamada05}
BaCu$_2$SiO$_6$,\cite{Jaime04,Sebastian05}
CsCuCl$_4$,\cite{Radu05,Sebastian06b} and
NiCl$_2$-4SC(NH$_2$)$_2$. \cite{Zapf06,Zvyagin07}

The bosons that condense in these systems are not the atoms, but
rather Cu or Ni spin degrees of freedom (also referred to as
triplons or magnons in the literature.) The BEC corresponds to the
onset of magnetic order induced by magnetic fields. Thus, the
tuning parameter for inducing condensation in spin systems is not
the temperature, but the magnetic field.

In this paper, we focus on the organic quantum magnet
NiCl$_2$-4SC(NH$_2$)$_2$ (DTN). It is the first compound based on
Ni (spin S = 1) instead of Cu (S = 1/2) and the first organic
compound in which BEC has been observed. \cite{Zapf06} We present
an extended introduction to the topic of Bose-Einstein
condensation in DTN and then discuss new magnetostriction data for
this compound at low temperatures.

\begin{figure}[tbp]
\epsfbox{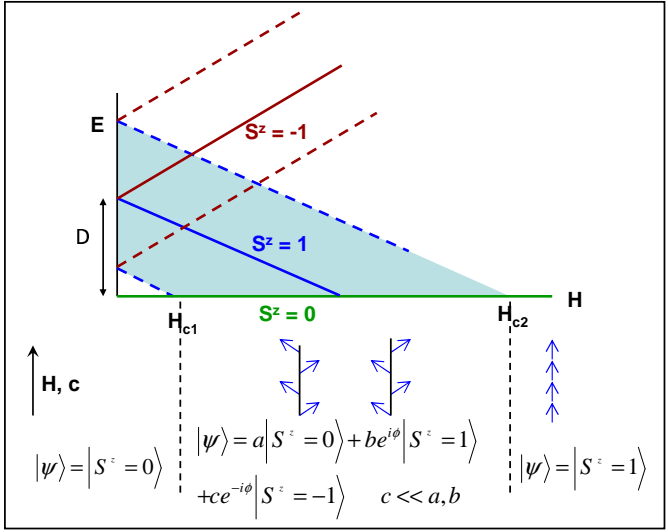} \caption{Energy level diagram of DTN.
The $S_z = \pm 1$ states are shown schematically as broad bands
due to antiferromagnetic dispersion. The spins (blue arrows) form
a disordered spin liquid with $S_z = 0$ below $H_{c1}$, then order
in a canted AFM structure between $H_{c1}$ and $H_{c2}$. Finally,
above $H_{c2}$ the spins polarize along the field direction. The
ground state wave function $|\psi>$ in these three regions is also
indicated.} \label{energy_levels}
\end{figure}

The energy level diagram of the Ni system differs from those of
the previously studied Cu systems. The Ni $S = 1$ spin triplet is
split by single-ion anisotropy into a $S_z = 0$ ground state and
$S_z = \pm 1$ excited states with an energy gap of $D \sim 10$ K.
\cite{PaduanFilho81,PaduanFilho04,Zvyagin07} The $S_z = \pm 1$
levels are dispersed by the antiferromagnetic coupling J, e.g. the
spins can have different energies depending on their orientation
with respect to each other. Thus these levels become broad bands,
as shown in Fig. \ref{energy_levels} where the bottom of the band
is the AFM $k = (\pi,\pi,\pi)$ vector. When a magnetic field is
applied along the tetragonal c-axis, the Zeeman effect lowers the
$S_z = 1$ level until the bottom of the band becomes degenerate
with the $S_z = 0$ ground state at $H_{c1}$ (see Fig.
\ref{energy_levels}). Between $H_{c1}$ and $H_{c2}$ the mean-field
ground state becomes a linear combination of all three energy
levels until finally at $H_{c2}$ the gap reopens now with the $S_z
= 1$ level as the ground state. \cite{Zapf06}

\epsfxsize=250pt
\begin{figure}[tbp]
\centering \epsfbox{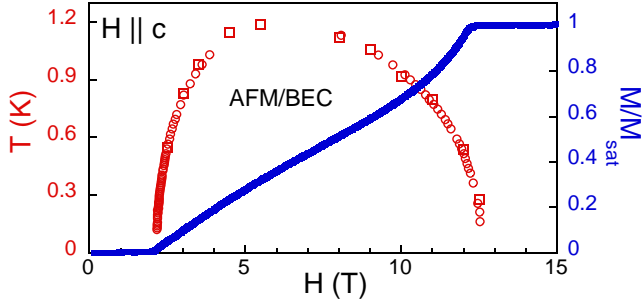}
\caption{Temperature $T$ - Magnetic field $H$ phase diagram for $H
|| c$ determined from specific heat and magnetocaloric data.
\cite{Zapf06} The magnetization vs field at 16 mK is overlayed
onto the phase diagram. \cite{PaduanFilho04}}
\label{introduction_phasediagram}
\end{figure}

\epsfxsize=250pt

At zero field, the gap between the $S_{z} = 0$  and the $S_{z} =
\pm 1 $ levels precludes any magnetic order. However, between
$H_{c1}$ and $H_{c2}$, XY antiferromagnetic order occurs at low
temperatures. The region of field-induced antiferromagnetic order
between $H_{c1}$ and $H_{c2}$ and below the maximum $T_N = 1.2$ K
is shown in the phase diagram in Fig.
\ref{introduction_phasediagram}. At $H_{c1}$, the spins order
antiferromagnetically in the a-b plane, perpendicular to the
applied field direction. As the field increases, the spins cant
along the field direction until finally at $H_{c2}$, the spins
polarize and the magnetization saturates. The magnetization along
the $c$ axis is shown overlaid on the phase diagram in Fig.
\ref{introduction_phasediagram}. We should note that this
description only holds for fields applied along the tetragonal
c-axis. For fields in the a-b plane, the $S_z = 0$ groundstate
mixes with a linear combination of the $S_z = \pm 1$ excited
states and the resulting new set of eigenstates move apart with
field, thus there is no level crossing and the system does not
order.

\begin{figure}[tbp]
\epsfbox{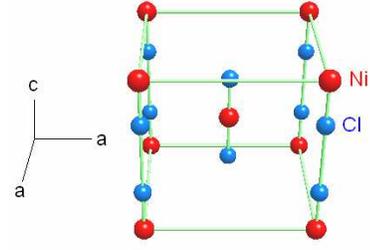} \caption{Unit cell of tetragonal
NiCl$_2$-4SC(NH$_2$)$_2$ showing Ni (red) and Cl (blue) atoms.
Remaining atoms have been omitted for clarity.}
\label{crystalstructure}
\end{figure}

Inelastic neutron scattering measurements in zero field
\cite{Zapf06} indicate that the antiferromagnetic coupling is
strongest along the Ni-Cl-Cl-Ni chains along the tetragonal
c-axis, with the AFM exchange parameter $J_c = 1.74(3)$ K. The
crystal structure of DTN is shown in Fig. \ref{crystalstructure}.
The coupling along the a-axis (and equivalently the b-axis), $J_a
= J_b = 0.17(1)$ K, is significantly weaker, and no couplings were
found to within the experimental resolution along the (1,1,1)
direction. Thus the magnetic structure of this compound consists
of semi-1D chains of Ni atoms that order three-dimensionally below
1.2 K in fields.

\epsfxsize=150pt

In DTN, the Bose statistics come about not because the spins are
$S = 1$. In fact, theoretically BEC could be observed in a purely
$S = 1/2$ magnetic system. Rather, the Bose statistics result from
the fact that only the spins play a role in the Hamiltonian, and
the atoms themselves are confined to lattice sites. The spin
operators on different sites commute with each other, leading to
Bose statistics.

 The spin $S=1$ system can be mapped into a gas of
semi-hard-core bosons \cite{Batista01,Batista04,Ng05} with no more
than two bosons per site. The $|S_z = 1>$, $|S_z=0>$ and $|S_z =
-1>$ states are mapped into the states with two, one and zero
bosons respectively. The transverse component of the Heisenberg
interaction is mapped into a hopping term while the longitudinal
or Ising component becomes a nearest-neighbor repulsion. The
single-ion anisotropy $D$ generates an on-site repulsion and the
magnetic field is mapped into a chemical potential. The resulting
Hamiltonian in the particle representation is an extended Hubbard
model for the semi-hard-core bosons. In the particle
representation, the phase for $H < H_{c1}$ corresponds to the Mott
insulator (one boson per site). When the chemical potential
(magnetic field) reaches a critical value $H = H_{c1}$, the
density $\rho$ of bosons (magnetization) starts to increase
continuously and the additional bosons relative to $\rho = 1$ form
a BEC state at low enough temperature. This state is the particle
version of the XY-antiferromagnet. When the chemical potential
reaches a second critical value $H = H_{c2}$, all the sites are
occupied by two bosons ($\rho = 2$) and the system enters a second
Mott insulating phase that is the particle version of the fully
saturated spin state.

The tetragonal crystal structure of DTN creates an approximate
uniaxial symmetry (U(1) symmetry) of the spin system. This
guarantees that the antiferromagnetic order for fields between
$H_{c1}$ and $H_{c2}$ is XY-like, meaning that there is no
energetically favorable direction for the spins to point in the
plane perpendicular to the applied field. The order parameter
therefore acquires not only a magnitude but also a phase and the
resulting Bose-Einstein Condensate exhibits spontaneous phase
coherence.

There are several caveats to the idea of BEC in quantum spin
systems. First of all, the uniaxial spin symmetry of the
Hamiltonian is approximate. The square lattice of the crystal
introduces anisotropy in the a-b plane as a result of spin-orbit
coupling and dipole-dipole interactions. \cite{Sebastian06a}
However, these effects generally occur at lower energy scales
(often $\mu$K) and can thus be neglected at the temperatures of
tens to hundreds of mK at which these quantum magnets are studied.
In particular, no significant Dzyaloshinskii-Moriya interactions
have been observed in electron-spin-resonance measurements of DTN.
\cite{Zvyagin07}

Thus, the Bose-Einstein condensation picture in quantum magnets is
valid for the temperatures at which these compounds are studied,
and more importantly it provides a way of understanding the
observed physical behavior. The thermal phase transition in this
system belongs to the $d = 3$ universality class of an XY
antiferromagnet, where $d$ is the number of spatial dimensions.
However, the field-induced quantum phase transition belongs to the
$D = 5$ ($d = 3$, $z = 2$) universality class where $z$ is the
dynamical exponent.

One of the key experimental signatures of the Bose-Einstein
Condensation quantum critical point is the temperature-dependence
of the critical field $H_{c1}$ near $T = 0$. For a 3-D BEC,
$H_{c1}(T) \propto T^{\alpha}$ where $\alpha = 3/2$. In contrast,
the prediction for an Ising magnet is $\alpha = 2$, and for a 2-D
BEC, $\alpha = 1$ ) (T corresponds to the
Berezinskii-Kosterlitz-Thouless transition temperature in this
case).  In DTN, $H_{c1}(T)$ as $T \rightarrow 0$ has been
investigated in detail by specific heat and magnetocaloric effect
measurements. \cite{Zapf06} These data show that the expected
power-law behavior $H_{c1}(T) - H_{c1}(0) \propto T^{3/2}$ for 3-D
BEC is approached as $T \rightarrow 0$. \cite{Zapf06}

In this work, we present magnetostriction measurements of DTN
performed in a dilution refrigerator in a 20 T magnet system. The
original motivation for these experiments, as mentioned in a
previous work, \cite{Zapf06} was to account for a discrepancy
between the observed and the predicted phase diagram. A
semi-classical spin-wave approach predicted that $H_{c2} = 10.8$
T, in contrast to the observed value of $H_{c2} = 12.6$ T. In
fact, it turns out that this discrepancy results from the presence
of strong quantum fluctuations that limit the validity of a
spin-wave approach. Taking fluctuations into account, the
calculated value of $H_{c2}$ is very close to the experimental
one. Quantum Monte Carlo calculations are able to model both the
$T-H$ phase diagram and the magnetization as a function of field
very well. \cite{Zvyagin07} The magnetostriction is therefore not
required to account for any deviations in the observed results.

Nevertheless, upon measuring magnetostriction in this compound we
have found significant field-induced lattice distortions in DTN,
and these indicate that magnetoelastic coupling plays a role in
this system. We find field-induced changes in the lattice
parameters up to 0.025\%, which is a relatively large effect. We
also map the magnetic phase diagram as a function of magnetic
field angle from the tetragonal c-axis.

\section{Results and Discussion}

Magnetostriction measurements of single crystals of DTN were
performed in a titanium dilatometer. \cite{Schmiedeshoff06} The
sample was mounted on a titanium base plate using Si vacuum
grease. Length changes were monitored capacitatively with a Be-Cu
spring-mounted titanium tip and the sample was protected from the
tip with a .003" Ti foil. The dilatometer was mounted in a plastic
rotator inside the $^3$He-$^4$He mixture of a top-loading dilution
refrigerator, in a 20 T superconducting magnet system at the
National High Magnetic Field Laboratory in Tallahassee, FL.
Temperature was monitored with a ruthenium oxide thermometer, and
capacitance changes were measured using a digital capacitance
bridge operating at 5 kHz (Andeen-Hagerling model 2700A). Data is
shown above $H = 1$ T, since flux jumps in the Nb$_3$Sn
superconducting magnet affect the data at low fields.

New magnetization data for DTN is also presented. The longitudinal
magnetization for $H || a$ was measured in a vibrating sample
magnetometer (VSM) at 0.5 K. \cite{Oliveira72}

\begin{figure}[tbp]
\epsfbox{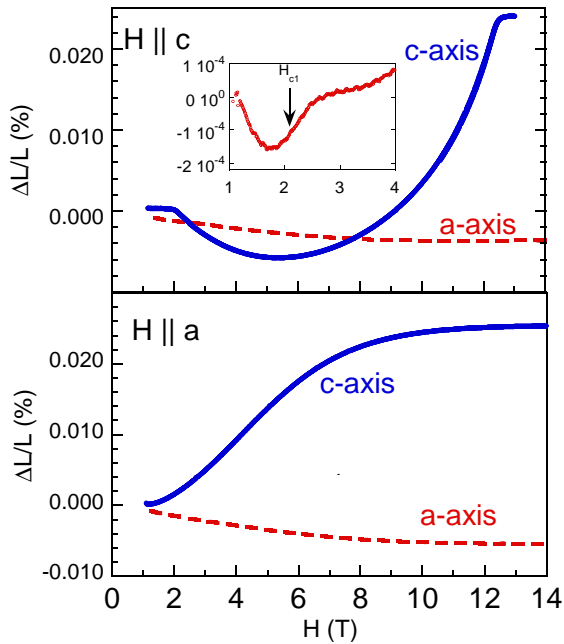} \caption{Normalized percentage
length change \%$\Delta L/L$ as a function of magnetic field
measured along the crystallographic c-axis (solid blue lines) and
a-axis (dashed red lines). Data is taken at 25 mK. The magnetic
field is applied along the c-axis (top), and a-axis (bottom). The
inset to the top figure is zoomed in on the feature at $H_{c1}$ in
\%$\Delta L_a/L_a$. A straight line has been subtracted from the
data for clarity. }\label{Magnetostriction}
\end{figure}

Magnetostriction data at 25 mK is shown in Fig.
\ref{Magnetostriction} for magnetic fields along the
crystallographic c- and a-axes (top and bottom figures,
respectively). For both field orientations, magnetostriction was
measured along the a- and c-axes of the crystal, e.g. both
perpendicular and parallel to the applied field.

\begin{figure}[tbp]
\epsfbox{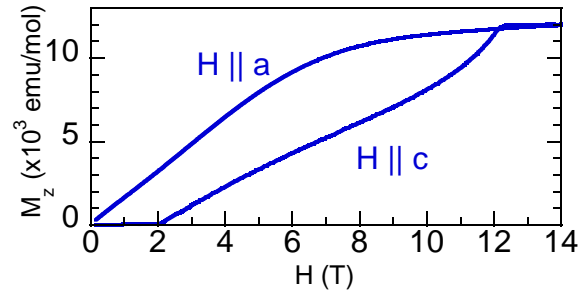} \caption{Longitudinal magnetization of
DTN for $H || c$ and $H || a$. The $H || c$ data is taken at 16 mK
in a force magnetometer, \cite{PaduanFilho04} and the $H || a$
data is taken in a VSM at $T = 0.5$ K.}\label{Magnetization}
\end{figure}

For $H || c$ (Fig. \ref{Magnetostriction} top,) changes in the
length of the sample along the c axis, $\Delta L_c$, clearly show
features at the magnetic phase transitions at $H_{c1}$ and
$H_{c2}$. $\Delta L_c$ is nearly constant below $H_{c1}$ and above
$H_{c2}$, but varies dramatically in the region of
antiferromagnetic order between $H_{c1}$ and $H_{c2}$. Between
$H_{c1} = 2.1$ T and $\sim 5$ T, $L_c$ shrinks, and between 5 T
and $H_{c2} = 12.5$ it expands, finally saturating at $H_{c2}$.
This behavior is rather startling since the magnetization for $H
|| c$ increases roughly linearly with field between $H_{c1}$ and
$H_{c2}$. The longitudinal magnetization for $H || c$ is shown for
comparison in Fig. \ref{Magnetization}. We interpret this
nonmonotonic behavior in $\Delta L_c$ as arising from competing
effects of antiferromagnetic and ferromagnetic components of the
canted order. At $H_{c1}$ the magnetic system orders purely
antiferromagnetically in the a-b plane. Thus, by shrinking the
distance between Ni atoms along the c-axis, the system can
increase $J_c$ and lower its magnetic energy. However, with
increasing field, the spins cant along the field direction,
increasing the ferromagnetic component. At some point it becomes
more energetically favorable for the system to expand the c-axis
thereby lowering the antiferromagnetic coupling $J_c$. Finally,
above $H_{c2}$ the spins are saturated along the applied field
direction and $L_c$ remains roughly constant. The magnitude of the
overall change in $L_c$ between $H_{c1}$ and $H_{c2}$ is $\sim
.025\%$. Such a distortion would be considered large in a metal
and would most likely correspond to a structural phase transition.
However, since DTN is a soft organic compound, a .025\% distortion
of the lattice parameters is reasonable.

 \epsfxsize=220pt

The $a$-axis magnetostriction $\Delta L_a$ for $H || c$, shown as
the red line in Fig. \ref{Magnetostriction} (bottom), is smaller
$\Delta L_c$ and $H_{c1}$ is barely distinguishable at 2.1 T. The
smaller magnetostriction along $a$ makes sense since $J_a << J_c$.
What is interesting is that $\Delta L_a$ doesn't follow the same
nonmonotonic behavior as $\Delta L_c$. One might expect at first
glance that $\Delta L_a$ should echo the same behavior as $\Delta
L_c$, just with a smaller magnitude. In reality, as the magnetic
field becomes larger, $\Delta L_a$ doesn't reverse directions like
$\Delta L_c$ does as the spins cant, but rather continues to
decrease until it saturates near $H_{c2}$. This odd behavior could
be due to the Poisson effect, e.g. volume-conserving forces
typically present in tetragonal crystals cause $L_a$ to contract
as $L_c$ expands. Thus at high fields, the $a$-axis lattice
parameter responds more strongly to the distortions in the
$c$-axis parameter than to magnetic forces from the
antiferromagnetic coupling $J_a$.

 \epsfxsize=220pt

For the other magnetic field direction, $H || a$, no magnetic
order has been observed, and the system behaves paramagnetically.
The magnetostriction shown in Fig. \ref{Magnetostriction} (top)
reflects this, and evolves smoothly with increasing magnetic
field, eventually saturating at high fields. Similar to the $H ||
c$ case, $L_c$ shows the strongest effect, and $L_a$ moves in the
opposite direction. The magnetization for $H || a$ is shown for
comparison in Fig. \ref{Magnetization}.

\epsfxsize=250pt
\begin{figure}[tbp]
\centering \epsfbox{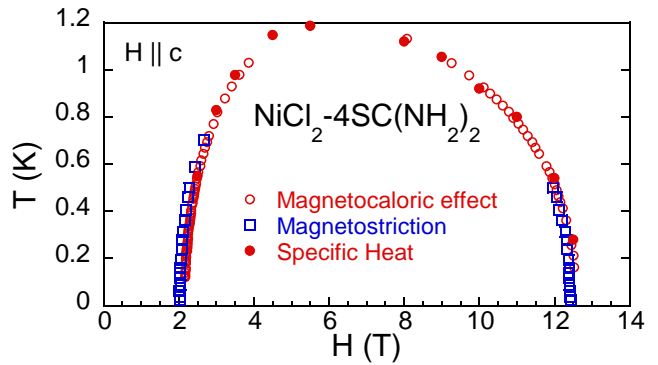} \caption{Temperature $T$ -
Magnetic field $H$ phase diagram for $H || c$ determined from
magnetostriction, with the phase diagram from specific heat and
magnetocaloric measurements shown for comparison \cite{Zapf06}.
The magnetostriction phase diagram (squares) is shifted to
slightly lower fields, which may result from a slight pressure
applied by the dilatometer.} \label{PhaseDiagram}
\end{figure}

The critical fields $H_{c1}$ and $H_{c2}$ can be extracted from
the magnetostriction data $\Delta L_c$ for $H || c$. We take the
peak in the second derivative of $\Delta L_c(H)$ to be the
antiferromagnetic transition. We have used this technique to map
the phase diagram as a function of temperature and field angle.
The resulting temperature-field phase diagram is shown in Fig.
\ref{PhaseDiagram}, along with that determined from specific heat
and magnetocaloric effect data for comparison.\cite{Zapf06} The
phase diagram from magnetostriction closely tracks that determined
from previous measurements, however, the phase transitions are
shifted to slightly lower fields. The discrepancy between the
phase diagrams cannot be accounted for by errors in the magnetic
field centering, or by rotational misalignments. We assume
therefore that this shift is due to pressure caused by the
spring-mounted dilatometer.

\epsfxsize=220pt
\begin{figure}[tbp]
\epsfbox{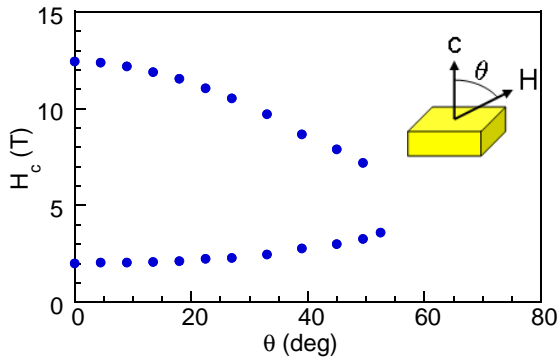} \caption{Critical fields $H_{c1}$ and
$H_{c2}$ at $T = 25$ mK as a function of field angle from the
tetragonal c-axis determined from magnetostriction data.}
\label{Angle}
\end{figure}

The critical fields $H_{c1}$ and $H_{c2}$ were also determined for
different field angles $\theta$ to the tetragonal c-axis. The
field-angle phase diagram is shown in Fig. \ref{Angle}. The region
of antiferromagnetic order between $H_{c1}$ and $H_{c2}$ shrinks
as $\theta$ increases, with antiferromagnetic order finally
vanishing above $\theta = 55 \deg$.

\section{Conclusions}

In conclusion, the compound NiCl$_2$-4SC(NH$_2$)$_2$ shows
significant magnetoelastic effects, which indicate a strong
coupling between the magnetic spin system and the lattice. As
discussed previously, the observed magnetostriction data can be
accounted for by magnetic forces resulting from the canted
antiferromagnetic order, and by the Poisson effect that conserves
the volume of the tetragonal crystal. The magnetostriction effects
along the crystallographic c-axis are so large that the
antiferromagnetic phase transition can be extracted from the data.
The phase diagram as a function of temperature, field, and angle
has been determined from the data, and agrees roughly with
specific heat and magnetocaloric effect measurements.

The original motivation for these measurements was to account for
a discrepancy between spin-wave theory prediction of $H_{c2} =
10.85$ T and the observed value of $H_{c2} = 12.6$ T.
\cite{Zapf06} The theory assumes constant values of $J_a$, $J_c$,
and $D$, whereas a magnetostriction effect would cause $J_a$,
$J_c$, and $D$ vary with field.

However, this large discrepancy in $H_{c2}$ has been mostly
resolved by taking into account spin fluctuations. Nevertheless,
small discrepancies between the theory and the experimental phase
diagram and the magnetization vs field data still exist.
\cite{Zvyagin07} These small discrepancies might still be
accounted for by the structural distortions of the sample with
field. The theoretical predictions for $H_{c1}$ and $H_{c2}$ are a
'best fit', obtained by varying the measured parameters $D$, $J_a$
and $J_c$ within their error bars. Thus both $H_{c1}$ and $H_{c2}$
show small discrepancies between theory and experiment. It is
possible that a fit that preferentially matches $H_{c1}$ would
result in a larger discrepancy in $H_{c2}$, which is in fact
accounted for by magnetostriction effects at high fields.

An important question that arises from these results is whether
the tetragonal symmetry of the crystal is preserved or whether the
magnetostriction effects induce a orthorhombic distortion. If the
tetragonal symmetry of the crystal is compromised in high fields
and the energy scale of these distortions is significant then the
Bose-Einstein Condensation description of the magnetic phase
transitions might no longer be valid. It would only be an accurate
description at $H_{c1}$ where the structural distortions go to
zero. In order to investigate the possibility of structural
distortions, resonant ultrasound measurements are in progress.

This work was supported by the DOE, the NSF, and Florida State
University through the National High Magnetic Field Laboratory.
V.S.Z. acknowledges support through the LANL Director-Funded
Postdoctoral program and A.P.F. acknowledges support from CNPq
(Conselho Nacional de Desenvolvimento Científico e Tecnológico,
Brazil).


\end{document}